\documentclass{iaus}
\usepackage{graphics}

  \checkfont{eurm10}
  \iffontfound
    \IfFileExists{upmath.sty}
      {\typeout{^^JFound AMS Euler Roman fonts on the system,
                   using the 'upmath' package.^^J}%
       \usepackage{upmath}}
      {\typeout{^^JFound AMS Euler Roman fonts on the system, but you
                   dont seem to have the}%
       \typeout{'upmath' package installed. iaus.cls can take advantage
                 of these fonts,^^Jif you use 'upmath' package.^^J}%
      }
  \else
  \fi

  \checkfont{msam10}
  \iffontfound
    \IfFileExists{amssymb.sty}
      {\typeout{^^JFound AMS Symbol fonts on the system, using the
                'amssymb' package.^^J}%
       \usepackage{amssymb}%
         \let\leq=\leqslant
         
      }{}
  \fi

  \IfFileExists{amsbsy.sty}
    {\typeout{^^JFound the 'amsbsy' package on the system, using it.^^J}%
     \usepackage{amsbsy}}
    {}

\newsavebox{\astrutbox}
\sbox{\astrutbox}{\rule[-5pt]{0pt}{20pt}}

\title[The role of interactions]{The role of interactions}

\author[H. R. Schmitt]%
{Henrique R. Schmitt$^1$}

\affiliation{$^1$National Radio Astronomy Observatory,
520 Edgemont Road, Charlottesville, VA22903, USA email:hschmitt@nrao.edu}

\pubyear{2004}
\volume{222}
\pagerange{1--8}
\date{?? and in revised form ??}
\jname{The Interplay among Black Holes, Stars and ISM \\in Galactic Nuclei}
\editors{T. Storchi-Bergmann, L. C. Ho \& H. R. Schmitt, eds.}
\begin{document}

\maketitle

\begin{abstract}
Interactions between galaxies are suggested to be a mechanism
responsible for feeding Active Galactic Nuclei (AGN). Theoretical models show
that interactions are an efficient way to drive gas from the galaxy to
the nucleus, however, the observational evidence on this subject is
controversial. Here we review results in this field, discuss possible
limitations of previous studies and the importance of dealing
with selection effects. We also show that there is no significant difference
in the percentage of low luminosity AGN and normal galaxies with companions,
and discuss possible interpretations of this result.
\end{abstract}

\firstsection 
\section{Introduction}

One of the major problems in the study of AGNs is understanding the
mechanisms responsible for making the gas lose large amounts of angular
momentum, move from the host galaxy toward the nucleus, and feed the
black hole. Among the mechanisms that have been proposed, the principal
ones are interactions, bars and nuclear spirals. The importance of
bars and nuclear spirals have been addressed in several contributions
to this conference (Combes, Crenshaw, Gadotti, Emsellem, Martini,
Maciejewski, among others), and here we will discuss the role of interaction.

Starting with Gunn (1979), who suggested that interactions can play an
important role in feeding AGNs, this mechanism has been studied by several
groups, both from the theoretical and observational point of view.
From the theoretical point of view, N-body simulations by Noguchi (1987),
Barnes \& Hernquist (1992), Byrd et al. (1986), Lin, Pringle \& Rees (1988),
Hernquist \& Mihos (1995), Mihos \& Hernquist (1994), Taniguchi \& Wada
(1996), among others, have shown that mergers, minor mergers and flybys can
be responsible for bringing gas from the disk to the nuclear region.
During this process the gas is shocked and compressed, leading to a period
of enhanced star formation before it can be accreted by the black hole.
This prediction is confirmed by the observation of higher levels of H$\alpha$,
infrared and radio emission in interacting galaxies (Keel et al. 1985;
Kennicutt et al. 1987). Further evidence comes from the fact that luminous
and ultraluminous infrared galaxies are closely related to interacting
systems (Sanders \& Mirabel 1996).

On the other hand, the observational evidence for the influence of interactions
in feeding AGNs is not very clear. In the case of high luminosity sources,
like Quasars, there is evidence that they are related to interacting
systems (Hutchings 1982; Heckman et al. 1983; Canalizo \& Stockton 2001).
However, in the case of lower luminosity sources, like Seyfert galaxies,
the situation is not so clear. There is no consensus on the importance of
interactions in feeding their nuclei, and whether these galaxies have an
excess of companions. The current results can be divided into 3 groups: those
with an excess of Seyfert galaxies with companions relative to normal
galaxies (e.g. Simkin, Su \& Schwarz 1980; Dahari 1984; Rafanelli,
Violato \& Baruffolo 1995; Laurikainen et al. 1994); those with no
difference between Seyferts and normal galaxies (Fuentes-Williams \&
Stocke 1988; Ulvestad \& Wilson 1984; De Robertis, Yee \& Hayhoe 1998);
and those with a higher percentage of Seyfert 2s than Seyfert 1s
with companions (Laurikainen \& Salo 1995, and Dultzin-Hacyan et al. 1999).

We believe that the contradictory results obtained by different papers are
most likely due to the way they selected their samples and control samples, as
pointed out by Heckman (1990). A major requirement for one to be able to do a
proper comparison between the percentage of AGN and normal galaxies with
companions, is to have a sample of active and control galaxies matched
by their host properties. Previous studies also had to deal with the fact
that they usually did not have information about the redshifts of nearby
galaxies projected on the sky, so they were not always able to know
if these galaxies were physically associated. This required the
application of statistical corrections for the number of background and
foreground objects projected around the galaxies being studied.

\section{Low Luminosity AGN}

In this section we present the results of a recent study of the percentage
of companions in galaxies with different activity types (Schmitt 2001),
where most of the selection effects, as well as other problems pointed out
above, were avoided. Instead of using the more traditional technique of
selecting a control sample of galaxies to match the AGN one, the approach
used in this paper was to employ a sample of galaxies selected by their host
properties, independent of the fact of being active or not. The sample
selected for this study was the Palomar survey (Ho, Filippenko \& Sargent
1997a), which comprises all the galaxies brighter than B$_T=12.5$ mag in the
northern hemisphere. This sample is ideal for this work because it includes
both the AGN and the control sample in itself.

Besides mitigating the sample selection problem, the Palomar sample presents
several additional advantages relative to other samples.
It provides homogeneous high-quality spectroscopic measurements
of emission line fluxes, and activity classification of all the galaxies.
This sample contains a large enough number of galaxies, with a large range
of parameters (e.g. host galaxy morphology and nuclear luminosity)
ensuring a significant comparison between the different activity types, and
allowing us to draw robust statistical conclusions. For a detailed discussion
of other advantages of this sample relative to other ones commonly used in the
literature, and for a discussion of possible selection biases which
affect them, see Ho \& Ulvestad (2001).

Starting with the original sample of 486 galaxies from Ho et al. (1997a),
we excluded local group galaxies and all those with B$_T>12.5$ mag,
thus reducing the total number of objects to 451. These galaxies were
divided in 5 groups, according to their nuclear activity type. The groups and
the number of galaxies in each one of them are: 46 Seyferts, 193 HII galaxies,
88 LINERs, 63 Transition objects and 61 Absorption line galaxies. We point out
that most of the HII galaxies in this sample have only quiescent star
formation. Also, we do not try to separate the Seyfert galaxies into
Seyfert 1s and Seyfert 2s.

The influence of interactions on AGN activity was studied using two
techniques. First we compared the environments where galaxies with different
activity types are found. This was done using the local galaxy
density parameter ($\rho_{gal}$), which was defined by Tully (1988) as the
density of galaxies brighter than M$_B=-16$ mag in the vicinity of
the object of interest. We found that the different activity types have
similar distributions of $\rho_{gal}$ values, indicating that there is
no significant difference in their environments (see the contribution by
Maia for results on a different sample).

\begin{figure}
\centering
\resizebox{13.5cm}{!}{\includegraphics{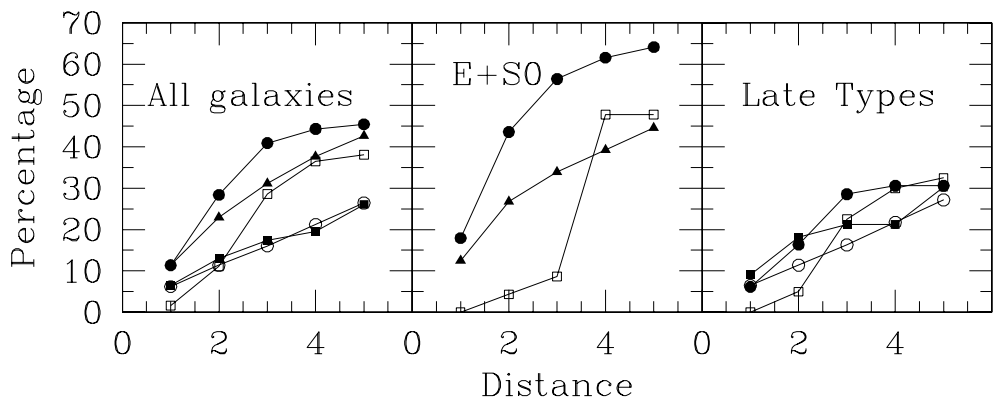}}
  \caption{Percentage of galaxies with companions as a function of the distance
between the two galaxies, in units of the diameter of the primary galaxy.
LINER's are represented by filled circles, Transition galaxies by open
squares, Absorption line galaxies by filled
triangles, Seyfert galaxies by filled squares and HII galaxies by open
circles. The left panel shows the results for all the galaxies, the middle
panel shows the results for Elliptical and S0 galaxies only, while the
right panel shows the results for late type galaxies only.}\label{fig1}
\end{figure}

The second technique consisted of searching for companions around the galaxies
in the Palomar sample. This was done using NED and the Digitized Sky Survey
images. The criteria used to determine if a galaxy has a companion
are similar to the ones used by Rafanelli et al. (1995), but modified
to allow a slightly larger search region. We determine that a galaxy is a
companion if its distance to the galaxy of interest, the primary galaxy,
is smaller than 5 times the diameter (D$_{25}$) of that galaxy, the
difference in brightness between them is smaller than 3 magnitudes
($|\Delta m|\leq$ 3 mag), and the difference in radial velocities is smaller
than 1000 km~s$^{-1}$ ($|\Delta V_{Rad}|\leq$ 1000 km~s$^{-1}$). The Palomar
sample presents a major advantage, relative to most samples, in the latter
criterion. Since redshift
surveys such as the CfA2 (Falco et al. 1999) are observing galaxies up
to $B\leq15.5$, we are able to obtain radial velocity information
even for the faintest possible companions of galaxies in our sample.

Considering that most of the galaxies in this sample have M$_B<-19$,
we can calculate that this technique is sensitive to companions as faint as
M$_B\sim-16$, which is similar to the absolute magnitude of the SMC
(M$_B=-16.3$). For more typical galaxies, with magnitudes around M$^*$
(M$_B^*=-20.3$), this technique is sensitive to companions slightly fainter
than the LMC (M$_B=-18.5$). However, we are not sensitive to lower luminosity
dwarf spheroidal and irregular galaxies, which were suggested by De Robertis
et al. (1998) as possible triggers of activity in Seyfert galaxies.

The results from the comparison of the percentage of galaxies with companions
among the different activity types are presented in Figure~\ref{fig1}.
We find in the left panel of this figure that, when all the galaxies in the
sample are considered, there is a similar percentage of Seyfert and HII
galaxies with companions. However, a surprising result arises when we compare
these percentages with those for LINER's, Transition and Absorption galaxies,
which present significantly larger percentages of companions, by a factor
of approximately two. A detailed statistical comparison between the different
activity types is presented in Schmitt (2001).

This result seems to contradict all previous results available in the
literature, which indicated that the percentage of Seyfert and HII galaxies
with companions was either similar to or higher than that of other galaxies.
However, this contradiction is solved
when we take into account the morphological types of the galaxies in our
sample. Ho et al. (1997b) showed that different activity types have
different distributions of morphological types. A particularly important
result of their analysis is the fact that most Absorption galaxies and
a significant percentage of LINERs and Transition galaxies are found in
Elliptical or S0 hosts. On the other hand, almost all HII galaxies 
and a high percentage of Seyferts have morphological types Sa or
later. Since the percentage of elliptical galaxies increases with the
density of the environment, what is known as the morphology-density effect
(Dressler 1980, Charlton et al. 1995; Budav\'ari et al. 2003),
one would expect to find a higher percentage of elliptical than spiral
galaxies with companions. As a consequence of this effect, the percentage of
LINER, Transition and Absorption galaxies with companions should on
average be higher than in the case of Seyfert and HII galaxies.

To address this issue we separate our sample into two groups, those with
morphological types E and S0, and those with morphological types Sa and later.
The results are presented in the middle and right panels of Figure~\ref{fig1},
respectively.
The middle panel shows the percentage of early type galaxies with companions,
as a function of the separation between the galaxies, where we can see that
LINER, Transition and Absorption line galaxies have similar percentages of
companions. This panel does not include Seyfert and HII galaxies because the
sample does not have many of these galaxies in Ellipticals and S0's hosts.
The comparison between late type galaxies shows that all activity classes have
similar percentages of galaxies with companions. Similar results were found
by Ho, Filippenko \& Sargent (2003), using only galaxies with morphological
types between Sab and Sbc. Comparing the percentage of early and late type
LINER's with companions we find a frequency 2 times higher in the former,
indicating that the contradictory results obtained using all galaxies in the
sample were due to the mismatch of morphological types of the different
activity classes.

Finally, we address whether there is a difference in the percentage of
Seyfert 1s and Seyfert 2s with companions. Laurikainen \& Salo (1995)
and Dultzin-Hacyan et al. (1999) found that, when compared to normal
galaxies, Seyfert 1s have a similar percentage of companions, while for
Seyfert 2s this percentage doubles. However, as discussed in Schmitt et al.
(2001), their results can be attributed to the way their samples were selected,
based on ultraviolet excess. Since the nucleus of Seyfert 2s is hidden from
direct view, the amount of ultraviolet emission reflected in our direction is
very small, indicating that the ultraviolet excess detected in their Seyfert 2s
is due to a source outside the torus, like a starburst. Since strong star
formation is related to interactions (see next section), and a large
percentage ($\sim$40\%) of Seyfert galaxies are known to have circumnuclear
star formation (Cid Fernandes et al. 2001), it is likely that their results
were influenced by selection effects. In fact, Schmitt et al. (2001) show
that, when we use a sample of Seyfert galaxies selected from an
isotropic property, 60$\mu$m luminosity, both Seyfert types have similar
percentages of galaxies with companions. More recently, Ho et al. (2003)
found a higher percentage of Seyfert 1s than Seyfert 2s with companions
in the Palomar sample. Nevertheless, they note that this result can be due
to the fact that some broad-lined Seyfert galaxies in their sample are
actually Seyfert 2s. Due to the high sensitivity of this survey, they
were able to detect the reflected broad lines in some Seyfert 2 galaxies
like in NGC1068, in direct light. Taking into account cases like this might
in fact remove the difference between the two Seyfert types.

\section{Higher Luminosities and Distances}

The results presented in the previous section indicate that there is no
significant difference in the percentage of normal galaxies and low luminosity
AGNs with companions, or in their environments. However, one might argue that
interactions may be more important for higher luminosity AGNs, or for sources
at higher redshifts. In the case of HII galaxies, we find that when we take
into account only the objects with L(H$\alpha$)$>10^{40}$ erg s$^{-1}$ in the
Palomar sample, 65\% of them have companions. For comparison, this percentage
drops to $\sim$30\% for lower luminosity sources. This result is consistent
with those from Kennicutt et al. (1987) and Keel et al. (1985), indicating
that interactions can be responsible for the enhancement of the luminosity
of these galaxies.

In the case of AGNs, studies by Brown et al. (2001), Miller et al. (2003)
and Grogin et al. (2003), based on strong AGNs at higher redshifts, found
that there is no difference in the environment or the number of companions
around these galaxies compared to normal ones. Nevertheless, a different
result was obtained by Kauffmann et al. (2004), based on a much larger
sample of AGNs extracted from the Sloan Survey. They found a higher fraction
of luminous AGN in lower density environments, which indicates that the
fraction of galaxies able to host strong AGN decreases in high-density
environments.

\section{Discussion and Future Work}

We saw in the previous sections that there is no difference in the percentage
of AGNs and normal galaxies with companions. Combining this fact with the
lack of an excess of AGNs, relative to normal galaxies, with bars or nuclear
spirals, one can interpret these results as evidence for these mechanisms
not being important in the AGN feeding process. However, a different
interpretation, in particular for the case of interactions, have been
proposed by several authors (Corbin 2000; Schmitt 2001; Storchi-Bergmann
et al. 2001; see also the
contributions by Martini, and Storchi-Bergmann et al.). These authors suggest
that these mechanisms play an important role in the feeding process, but
there is a delay between the time when the mechanism is active and when the
gas is being accreted by the nuclear black hole, indicating that the nuclear
activity may be cyclic and the AGNs evolve with time.

Support for this scenario comes from several sources. The results
obtained from N-body simulations (e.g. Byrd et al. 1986; Hernquist \& Mihos
1995) indicate that an interaction first moves gas toward the nucleus, where
its temperature and density increases and a period of strong star formation
happens. Later on this process the gas loses more angular momentum and is
accreted by the black hole, with the galaxy being detect as an AGN.
Taking into account the
fact that all galaxies with bulges are believed to have massive black holes
in their nuclei (Magorrian et al. 1998, Gebhardt et al. 2000, Ferrarese \&
Merritt 2000) it is reasonable to assume that when an interaction happens
they can pass through different periods of activity. First one would detect
an HII like nucleus, which could later evolve into a period of Seyfert
activity, when the gas is being accreted by the black hole at a high rate,
and finally become a LINER or transition galaxy, when the amount of gas
available to feed the nucleus is small. Observational evidence in favor of
this scenario comes from the detection of recent star formation, with ages
around 100~Myr, in the nuclear region of at least 40\% of Seyfert and
LINER/HII galaxies (Cid Fernandes et al. 2001, and this conference). 

The importance of interactions for the proposed scenario can be tested
observationaly. A relatively easy test is to compare
the asymmetry indices (Conselice 1997) of Seyfert and normal galaxies. If
interactions play an important role in feeding AGNs, one would expect larger
asymmetries in Seyfert galaxies compared to normal ones, or an evolution of
this index from one activity type to another. Similarly, it is also possible
to use HI~21~cm to detect the signature of interactions
no longer detectable in optical images. The contribution by Lim
et al. shows the results of such a study, where they find that Seyfert galaxies
present much more distorted and extended HI emission than normal galaxies.
It would be important to extend this study to a larger, well defined sample.

\begin{acknowledgments}
The National Radio Astronomy Observatory is a facility of
the National Science Foundation, operated under cooperative agreement by
Associated Universities, Inc.
\end{acknowledgments}

\end{document}